\documentstyle[epsf]{mn}

\newif\ifAMStwofonts

\ifoldfss
  \ifCUPmtlplainloaded \else
    \NewTextAlphabet{textbfit} {cmbxti10} {}
    \NewTextAlphabet{textbfss} {cmssbx10} {}
    \NewMathAlphabet{mathbfit} {cmbxti10} {} 
    \NewMathAlphabet{mathbfss} {cmssbx10} {} 
  \fi
  \ifAMStwofonts
    \ifCUPmtlplainloaded \else
      \NewSymbolFont{upmath} {eurm10}
      \NewSymbolFont{AMSa} {msam10}
      \NewMathSymbol{\upi}     {0}{upmath}{19}
      \NewMathSymbol{\umu}     {0}{upmath}{16}
      \NewMathSymbol{\upartial}{0}{upmath}{40}
      \NewMathSymbol{\leqslant}{3}{AMSa}{36}
      \NewMathSymbol{\geqslant}{3}{AMSa}{3E}

    \fi
  \fi
\fi 

\ifnfssone
  \newmathalphabet{\mathit}
  \addtoversion{normal}{\mathit}{cmr}{m}{it}
  \addtoversion{bold}{\mathit}{cmr}{bx}{it}
  \newmathalphabet{\mathbfit} 
  \addtoversion{normal}{\mathbfit}{cmr}{bx}{it}
  \addtoversion{bold}{\mathbfit}{cmr}{bx}{it}
  \newmathalphabet{\mathbfss} 
  \addtoversion{normal}{\mathbfss}{cmss}{bx}{n}
  \addtoversion{bold}{\mathbfss}{cmss}{bx}{n}
  \ifAMStwofonts
    \ifCUPmtlplainloaded \else
      \UseAMStwoboldmath
      \makeatletter
      \new@mathgroup\upmath@group
      \define@mathgroup\mv@normal\upmath@group{eur}{m}{n}
      \define@mathgroup\mv@bold\upmath@group{eur}{b}{n}
      \edef\UPM{\hexnumber\upmath@group}
      \new@mathgroup\amsa@group
      \define@mathgroup\mv@normal\amsa@group{msa}{m}{n}
      \define@mathgroup\mv@bold\amsa@group{msa}{m}{n}
      \edef\AMSa{\hexnumber\amsa@group}
      \makeatother
      \mathchardef\upi="0\UPM19
      \mathchardef\umu="0\UPM16
      \mathchardef\upartial="0\UPM40
      \mathchardef\leqslant="3\AMSa36
      \mathchardef\geqslant="3\AMSa3E
    \fi
  \fi
\fi 

\ifnfsstwo
  \DeclareMathAlphabet{\mathbfit}{OT1}{cmr}{bx}{it}
  \SetMathAlphabet\mathbfit{bold}{OT1}{cmr}{bx}{it}
  \DeclareMathAlphabet{\mathbfss}{OT1}{cmss}{bx}{n}
  \SetMathAlphabet\mathbfss{bold}{OT1}{cmss}{bx}{n}
  \ifAMStwofonts
    \ifCUPmtlplainloaded \else
      \DeclareSymbolFont{UPM}{U}{eur}{m}{n}
      \SetSymbolFont{UPM}{bold}{U}{eur}{b}{n}
      \DeclareSymbolFont{AMSa}{U}{msa}{m}{n}
      \DeclareMathSymbol{\upi}{0}{UPM}{"19}
      \DeclareMathSymbol{\umu}{0}{UPM}{"16}
      \DeclareMathSymbol{\upartial}{0}{UPM}{"40}
      \DeclareMathSymbol{\leqslant}{3}{AMSa}{"36}
      \DeclareMathSymbol{\geqslant}{3}{AMSa}{"3E}
    \fi
  \fi
\fi 

\ifCUPmtlplainloaded \else
  \ifAMStwofonts \else 
    \def\upi{\pi}
    \def\umu{\mu}
    \def\upartial{\partial}
  \fi
\fi

\newcommand{\Ell}{E_\parallel}      
\newcommand{\rhoGJ}{\rho_{{\rm GJ}}}  
\newcommand{\sgT}{\sigma_{\rm T}}  
\newcommand{\sgP}{\sigma_{\rm p}}  
\newcommand{\rlc}{\varpi_{\rm LC}} 
\newcommand{\Rc}{R_{\rm C}}        
\newcommand{\Ex}{\epsilon_{\rm x}} 
\newcommand{\Eg}{\epsilon_\gamma}  
\newcommand{\inc}{\alpha_{\rm i}}  
\newcommand{\figtwo} {\rm figure~3}   
\newcommand{\figfive}{\rm figure~7}   

\title[One-dimensional Electric Field Structure of an Outer Gap 
       Accelerator -- III]
      {One-dimensional Electric Field Structure of an Outer Gap 
       Accelerator -- III. 
       Location of the Gap and the Gamma-ray Spectrum}
\author[K. Hirotani and S. Shibata]
       {K. Hirotani${}^1$ and S. Shibata${}^2$ \\
        ${}^1$ National Astronomical Observatory,
	Osawa, Mitaka, Tokyo 181-8588, Japan \ 
        Email: hirotani@hotaka.mtk.nao.ac.jp\\
        ${}^2$ Department of Physics, Yamagata University,
	Yamagata 990-8560, Japan \ 
        Email: shibata@sci.kj.yamagata-u.ac.jp}
\date{Accepted  .
      Received 2000 November;
      in original form 2000 November}

\pagerange{\pageref{firstpage}--\pageref{lastpage}}
\pubyear{2001}

\begin{document}

\maketitle

\label{firstpage}

\begin{abstract}
We investigate a stationary particle acceleration zone
in the outer magnetosphere of a spinning neutron star.
The charge depletion due to a global current
causes a large electric field along the magnetic field lines.
Migratory electrons and/or positrons are accelerated by this field
to radiate curvature gamma-rays, 
some of which collide with the X-rays to materialize as pairs in the gap.
As a result of this pair--production cascade, 
the replenished charges partially screen the electric field, 
which is self-consistently solved together with the distribution
of particles and gamma-rays.
If no current is injected at neither of the boundaries of the accelerator,
the gap is located around the so-called null surface,
where the local Goldreich-Julian charge density vanishes.
However, we find that the gap position shifts outwards (or inwards)
when particles are injected at the inner (or outer) boundary.
We apply the theory to the nine pulsars of which X-ray fields are known 
from observations.
We show that the gap should be located near to or outside of 
the null surface for the Vela pulsar and PSR~B1951+32,
so that their expected GeV spectrum may be consistent with observations.
We then demonstrate that the intrinsically large TeV flux 
from the outer gap of PSR~B0540--69 is absorbed 
by the magnetospheric infrared photons to be undetectable.
We also point out that the electrodynamic structure and the resultant
GeV emission properties of millisecond pulsars are similar to 
young pulsars.
\end{abstract}

\begin{keywords}
gamma-rays: observation -- gamma-rays: theory -- magnetic field 
-- pulsars: individual 
(B0540--69, B1055--52, B1509--58, B1951+32, 
 Geminga, J0437--4715, J0822--4300, J1617--5055, Vela) 
\end{keywords}

\section{Introduction}

The EGRET experiment on the Compton Gamma Ray Observatory
has detected pulsed signals from seven rotation-powered pulsars
(for Crab, Nolan et al. 1993, Fierro et al. 1998;
 for Vela, Kanbach et al. 1994, Fierro et al. 1998;
 for Geminga, Mayer-Hassel Wander et al. 1994, Fierro et al. 1998; 
 for PSR B1706--44, Thompson et al. 1996;
 for PSR B1046--58, Kaspi at al. 2000;
 for PSR B1055--52, Thompson et al. 1999;
 for PSR B1951+32, Ramanamurthy et al. 1995).
The modulation of the $\gamma$-ray light curves at GeV energies 
testifies to the production of $\gamma$-ray radiation in the pulsar 
magnetospheres either at the polar cap 
(Harding, Tademaru, \& Esposito 1978; Daugherty \& Harding 1982, 1996;
 Dermer \& Sturner 1994; Sturner, Dermer, \& Michel 1995;
 Shibata, Miyazaki, \& Takahara 1998),
or at the vacuum gaps in the outer magnetosphere
(Cheng, Ho, \& Ruderman 1986a,b, hereafter CHR;
 Chiang \& Romani 1992, 1994; Romani and Yadigaroglu 1995;
 Romani 1996; Zhang \& Cheng 1997, hereafter ZC97).
Effective $\gamma$-ray production in a pulsar magnetosphere
may be extended to the very high energy (VHE) region above 
100 GeV as well;
however, the predictions of fluxes by the current models of 
$\gamma$-ray pulsars are not sufficiently conclusive
(e.g., Cheng 1994).
Whether or not the spectra of $\gamma$-ray pulsars continue up to the
VHE region is a question which remains one of the 
interesting issues of high-energy astrophysics.

In the VHE region,
positive detections of radiation at a high confidence 
level have been reported from the direction of the 
Crab, B1706--44, and Vela pulsars 
(Nel et al. 1993; Edwards et al. 1994;
 Yoshikoshi et al. 1997; 
 see also Kifune 1996 for a review),
by virtue of the technique of imaging Cerenkov light
from extensive air showers.
However, as for {\it pulsed} TeV radiation,
only the upper limits have been, as a rule, obtained from these pulsars
(see the references cited just above). 
If the VHE emission originates the pulsar magnetosphere,
rather than the extended nebula,
a significant fraction of them can be expected to show pulsation.
Therefore, the lack of {\it pulsed} TeV emissions provides a
severe constraint on the modeling of particle acceleration zones
in a pulsar magnetosphere.

In fact, in CHR picture,
the magnetosphere should be optically thick for pair--production
in order to reduce the TeV flux to an unobserved level 
by absorption.
This in turn requires very high luminosities of tertiary photons
in the infrared energy range.
However, the required fluxes are generally orders of magnitude
larger than the observed values (Usov 1994).
We are therefore motivated by the need to contrive an outer--gap model
which produces less TeV emission with a moderate infrared luminosity.

High-energy emission from a pulsar magnetosphere,
in fact, crucially depends on the acceleration electric field, 
$\Ell$, along the magnetic field lines.
It was Hirotani and Shibata (1999a,b,c; hereafter Papers I, II, III),
and Hirotani (2000a,b,c; hereafter Papers IV, V, VI)
who first solved the spatial distribution of $\Ell$ 
together with particle and $\gamma$-ray distribution functions.
They considered a pair-production cascade 
in a neutron star magnetosphere by solving the 
Vlasov equations (see also Beskin et al. 1992).

\begin{figure} 
\centerline{ \epsfxsize=9cm \epsfbox[0 70 500 400]{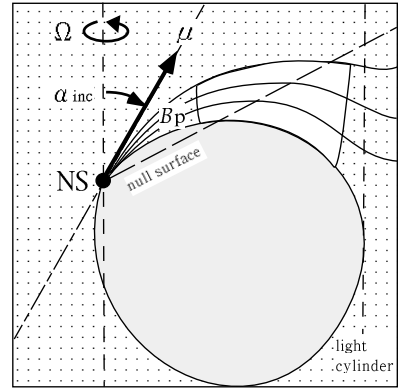} } 
\caption{\label{fig:oblique} 
Side view of a hypothetical outer-magnetospheric gap.
On the null surface, the Goldreich-Julian charge density vanishes.
        }
\end{figure}

By this method, 
they explicitly solved the gap width along the field lines, 
$\Ell$,
particle densities, and 
the $\gamma$-ray distribution functions.
They further demonstrated that 
a stationary gap is formed around the null surface at which the 
local Goldreich--Julian charge density, 
\begin{equation}
  \rho_{\rm GJ}= \frac{\Omega B_z}{2\pi c},
  \label{eq:def_rhoGJ}
\end{equation}
vanishes (fig.~\ref{fig:oblique}),
where $B_z$ is the component of the magnetic field along 
the rotation axis,
$\Omega$ refers to the angular frequency of the neutron star,
and $c$ is the speed of light.
Equation (\ref{eq:def_rhoGJ}) is valid unless the gap is 
located close to the light cylinder,
of which distance from the rotation axis is
given by 
\begin{equation}
  \rlc=3 \times 10^{8} \Omega_2{}^{-1} \mbox{cm},
  \label{eq:def_rlc}
\end{equation}
where $\Omega_2 \equiv \Omega/(100 \, \mbox{rad s}^{-1})$.
In this paper, 
we develop the method presented in Paper VI as follows: \\
$\bullet$ \ 
We compute the curvature radius, $\Rc$, at each point along the
magnetic field line, 
rather than assuming $\Rc= 0.5 \rlc$ throughout the gap.
\\
$\bullet$ \ 
We partially take the effect of the unsaturated particles
into account (eq.[\ref{eq:terminal}]).
\\
$\bullet$ \ 
We investigate the gap structure when particles flow
into the gap from the inner or the outer boundaries.
\\
$\bullet$ \ 
We compute the explicit spectra of TeV emission due to
inverse Compton scatterings,
rather than estimating only their upper limits.

In the next two sections, we describe the physical processes of pair 
production cascade in the outer magnetosphere of a pulsar.
We then apply the theory to individual pulsars 
and present the expected GeV and TeV spectra for various
boundary conditions on the current densities in \S~\ref{sec:appl}.
In the final section, we compare the results with previous works.

\section{Basic Equations and Boundary Conditions}
\label{sec:basicEQ}

We first reduce the Poisson equation for the electric potential
into a one-dimensional form in \S~\ref{sec:Poisson}.
We then give the Boltzmann equations of the particles and 
the $\gamma$-rays in \S\S~\ref{sec:contEQ} and \ref{sec:Boltzmann_gamma},
and describe the Lorentz factors and the X-ray fields in
\S\S~\ref{sec:terminal} and \ref{sec:X-ray}.
We finally impose suitable boundary conditions 
in \S~\ref{sec:BD}.

\subsection{Reduction of the Poisson Equation}
\label{sec:Poisson}

\begin{figure} 
\centerline{ \epsfxsize=9cm \epsfbox[0 110 450 300]{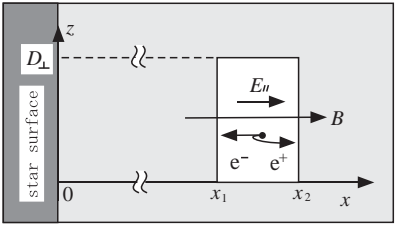} } 
\caption{\label{fig:rectil} 
Rectilinear approximation to the outer gap.
The $y$-axis, which designates the azimuth, is not depicted to
avoid complication.
This approximation is valid unless the gap width, $W= s_2 -s_1$,
becomes a good fraction of $\rlc$.
        }
\end{figure}

In the Poisson equation for the electrostatic potential $\Phi$,
the special relativistic effects appear in the higher orders
of $(\Omega \varpi / c)^2$,
where 
$\varpi$ indicates the distance of the point from the rotational axis
(Shibata 1995).
As the first-order approximation,
we neglect such terms.
To simplify the geometry, we further introduce a rectilinear approximation.
Unless the gap width, $W$, along the field lines becomes a good fraction
of $\rlc$,
we can approximate the magnetic field lines as straight lines
parallel to the $s$-axis (figure \ref{fig:rectil}).
Here, $s$ is an outwardly increasing coordinate along a
magnetic field line,
while $y$ designates the azimuth.
Assuming that typical transfield thickness of the gap, $D_\perp$, 
is greater than $W$, we can Fourier analyze in the transfield direction
to rewrite the Poisson equation into (eq.~[3] in Paper VI) 
\begin{equation}
  -\frac{d^2}{ds^2} \Phi 
    = -\frac{\Phi}{D_\perp{}^2}
      +4 \pi e \left( N_+ -N_- -\frac{\rhoGJ}{e} \right),
  \label{eq:Poisson_1}
\end{equation}
where $N_+$ and $N_-$ refer to the positronic and electronic densities,
respectively,
$e$ the magnitude of the charge on the electron,
and $s$ the length along the last-open fieldline.
In what follows, we define  
the origin of $s$ as the neutron star surface. 

It is convenient to non-dimensionalize the length scales 
by a typical Debey scale length $c/\omega_{\rm p}$, where
\begin{equation}
  \omega_{\rm p} = \sqrt{ \frac{4\pi e^2}{m_{\rm e}}
	                  \frac{\Omega B_{\rm cnt}}{2\pi ce} }
	         = 1.875 \times 10^7 \Omega_2{}^{1/2} B_5{}^{1/2}
		   \mbox{rad s}^{-1};
  \label{eq:def-omegap}
\end{equation}
$B_5$ represents the magnetic field strength 
at the gap center in $10^5$ G unit.
The dimensionless coordinate variable then becomes
\begin{equation}
  \xi \equiv (\omega_{\rm p}/c) s
      = 6.25 \times 10^{-4} \Omega_2{}^{1/2} B_5{}^{1/2} s .
  \label{eq:def-xi}
\end{equation}
By using such dimensionless quantities, we can rewrite
the Poisson equation into
\begin{equation}
  E_\parallel = -\frac{d\varphi}{d\xi},
  \label{eq:basic-1}
\end{equation}
\begin{equation}
  \frac{dE_\parallel}{d\xi}
  = - \frac{\varphi}{\Delta_\perp{}^2} +
      \frac{B(\xi)}{B_{\rm cnt}} \left[ n_+(\xi) - n_-(\xi) \right]
    + \frac{B_z(\xi)}{B_{\rm cnt}}
  \label{eq:basic-2}
\end{equation}
where $ \Delta_\perp \equiv (\omega_{\rm p}/c) D_\perp$ and 
$ \varphi(\xi) \equiv e\Phi(x)/(m_{\rm e}c^2)$;
the particle densities are normalized in terms of a
typical value of the Goldreich--Julian density as
\begin{equation}
  n_\pm(\xi) \equiv 
    \frac{2\pi ce}{\Omega} \frac{N_\pm (x)}{B(x)}.
  \label{eq:def-n}
\end{equation}
$B_{\rm cnt} \equiv B(\xi_{\rm cnt})$ refers to the magnetic field strength
at the gap center, $\xi= \xi_{\rm cnt}$.
In the non-relativistic limit, 
the local Goldreich--Julian charge density is given by
\begin{equation}
 \frac{B_z(\xi)}{B_{\rm cnt}}
    = \frac{ 2\cos\theta \cos(\theta-\inc)
             -\sin\theta \sin(\theta-\inc) }
           {\sqrt{1+3\cos^2\theta_{\rm cnt}}},
  \label{eq:rhoGJ_1}
\end{equation}
where $\inc$ refers to the inclination angle of the magnetic 
moment, $\theta$ and $\theta_{\rm cnt}$ the colatitude angle 
of the position at which $\rhoGJ$ is measured and that at the gap center,
respectively.
Neglecting the general relativistic effect near to the 
star surface, we can implicitly solve $\theta$ in terms of $s$ from
\begin{equation}
  \frac{s}{\rlc}
     = \int^\theta_{\theta_\ast}
       \frac{\sin(\theta-\inc)\sqrt{1+3\cos^2(\theta-\inc)}}
            {\sin\theta_{\rm LC}\sin^2(\theta_{\rm LC}-\inc)}
       d\theta,
  \label{eq:arc}
\end{equation}
where $\theta_\ast$ is the colatitude angle of the last-open
fieldline at the star surface; 
it is given by
\begin{equation}
  \frac{\sin^2(\theta_\ast-\inc)}{r_\ast}
    = \frac{\sin^2(\theta_{\rm LC}-\inc)}
           {\rlc / \sin \theta_{\rm LC}},
  \label{eq:r_ast}
\end{equation}
where $r_\ast$ refers to the neutron-star radius,
and $\theta_{\rm LC}$ the colatitude angle of the 
intersection of the last-open fieldline and the light cylinder.
We can solve $\theta_{\rm LC}$ for a given $\inc$ from
\begin{equation}
  2 \cot(\theta_{\rm LC}-\inc) \sin\theta_{\rm LC}
    + \cos\theta_{\rm LC} = 0.
  \label{eq:thLC}
\end{equation}

\subsection{Particle Continuity Equation}
\label{sec:contEQ}

Let us now consider the continuity equations of the particles.
It should be noted that almost all the particles migrate with the 
large, saturated Lorentz factor.
We therefore assume here that both electrostatic and 
curvature-radiation-reaction forces cancel out each other
to obtain the following continuity equations
\begin{equation}
  \pm B \frac{d}{ds}\left( \frac{N_\pm}{B} \right)
  = \frac{1}{c} \int_{0}^\infty d\epsilon_\gamma \, 
    [ \eta_{\rm p+} G_+   +\eta_{\rm p-} G_- ],
  \label{eq:cont-eq}
\end{equation}
where $G_\pm(x,\epsilon_\gamma)$ are the distribution function of
$\gamma$-ray photons having momentum $\pm m_{\rm e}c \epsilon_\gamma$ 
along the poloidal field line.
Since the electric field is assumed to be positive in the gap,
$e^+$'s (or $e^-$'s) migrate outwards (or inwards).
The pair production redistribution
function for outwardly (or inwardly) propagating $\gamma$-rays
are written as $\eta_{\rm p+}$ (or $\eta_{\rm p-}$) 
and are defined by
\begin{equation}
  \eta_{{\rm p}\pm}(\Eg)
  = (1-\mu_{\rm c}) \frac{c}{\omega_{\rm p}}
     \int_{\epsilon_{\rm th}}^\infty d\epsilon_{\rm x}
     \frac{dN_{\rm x}}{d\Ex} 
     \sgP(\Eg,\Ex,\mu_{\rm c}),
  \label{eq:def_etap_0}
\end{equation}
where $\sgP$ is the pair-production cross section and
$\cos^{-1}\mu_{\rm c}$ refers to the collision angle between 
outwardly (or inwardly) propagating $\gamma$-rays
and the X-ray photon with energy $m_{\rm e}c^2 \Ex$;
X-ray number density between  dimensionless energies
$\Ex$ and $\Ex+d\Ex$,
is integrated over $\Ex$ from
the threshold energy
$\epsilon_{\rm th} \equiv 2(1-\mu_{\rm c})^{-1} \Eg^{-1}$
to the infinity.
The explicit expression of $\sgP$ is given by 
(Berestetskii et al. 1989)
\begin{eqnarray}
  & & \sgP(\Eg,\Ex,\mu_{\rm c})
  \nonumber \\
  &\equiv& \frac{3}{16} \sgT
    (1-v^{2}) \left[ (3-v^4) \ln \frac{1+v}{1-v} -2v(2-v^{2}) \right],
  \nonumber
  \label{eq:def-sgP}
\end{eqnarray}
\begin{equation}   
  v(\Eg, \Ex, \mu_{\rm c}) 
     \equiv \sqrt{ 1 - \frac{2}{1-\mu_{\rm c}} \frac{1}{\Eg\Ex} },
  \label{eq:def-v}
\end{equation} 
where $\sgT$ is the Thomson cross section,
and $\Ex \equiv E_{\rm x}/m_{\rm e}c^2$
refers to the dimensionless energy of an X-ray photon.

To evaluate $\mu_{\rm c}$ between the surface X-rays and the $\gamma$-rays,
we must consider $\gamma$-ray's toroidal momenta due to an aberration
of light.
It should be noted that a $\gamma$-ray photon propagates 
in the instantaneous direction of the particle motion 
at the time of the emission.
The toroidal velocity of a particle at the gap center 
($r_{\rm cnt}$,$\theta_{\rm cnt}$)
becomes
\begin{equation}
  v_\phi= r_{\rm cnt} \sin\theta_{\rm cnt} \Omega
         +\kappa B_\phi
         -\Ell B_\phi \frac{c}{B^2},
  \label{eq:toroidal}
\end{equation}
where $\kappa$ is a constant.
On the right-hand side, the first term is due to corotation,
while second one due to magnetic bending.
Since $\Ell$ arises in the gap, the corresponding drift velocity
appears as the third term.
Unless the gap is located close to the light cylinder, 
we can neglect the terms containing $B_\phi$ as a first--order 
approximation.
We thus have $v_\phi=r_{\rm cnt} \sin\theta_{\rm cnt} \Omega$.
Since the three-dimensional particle velocity is virtually $c$,
the angle between the particle motion and the poloidal plane
becomes 
\begin{equation}
  \phi_{\rm ab}
     = \sin^{-1} \left( \frac{r_{\rm cnt} \sin\theta_{\rm cnt} \Omega}{c}
                 \right)
     = \sin^{-1} \left( \frac{r_{\rm cnt} \sin\theta_{\rm cnt}}{\rlc}
                 \right).
  \label{eq:abb}
\end{equation}
We evaluate ($r_{\rm cnt}$,$\theta_{\rm cnt}$) along a Newtonian dipole.
Using $\phi_{\rm ab}$, we can compute $\mu_{\rm c}$ as
\begin{equation}
  \mu_{\rm c}= \cos\phi_{\rm ab} \cos\theta_{\rm pol},
  \label{eq:def_coll_surface}
\end{equation}
where $\theta_{\rm pol}$ is the collision angle between 
the surface X-rays and the curvature $\gamma$-rays,
that is, the angle between the two vectors
($r_{\rm cnt}$,$\theta_{\rm cnt}$) and 
($B^r$,$B^\theta$) at the gap center.
Therefore, the collision approaches head-on (or tail-on) 
for inwardly (or outwardly) propagating $\gamma$-rays, 
as the gap shifts towards the star.

For magnetospheric, power-law X-rays, as assume
\begin{equation} 
  \mu_{\rm c}= \cos(W/\rlc),
  \label{eq:def_coll_power}
\end{equation}
for both inwardly and outwardly propagating $\gamma$-rays.
Aberration of light is not important for this component,
because both the X-rays and the $\gamma$-rays are emitted
nearly at the same place, provided that $W \ll \rlc$ holds.

Let us introduce the following dimensionless $\gamma$-ray 
densities in the dimensionless energy interval
between $\beta_{i-1}$ and $\beta_i$ such that
\begin{equation}
  g_\pm{}^i(\xi) \equiv 
    \frac{2\pi ce}{\Omega B_{\rm cnt}}
    \int_{\beta_{i-1}}^{\beta_i} d\epsilon_\gamma G_\pm(x,\epsilon_\gamma).
  \label{eq:def-g}
\end{equation}
In this paper, we set $\beta_0=10^2$;
this implies that the lowest $\gamma$-ray energy considered is
$m_{\rm e}c^2= 51.1$~MeV.
To cover a wide range of $\gamma$-ray energies, 
we divide the $\gamma$-ray spectra into $9$ energy bins 
and put
$\beta_1= 3.162 \times 10^2$, 
$\beta_2= 10^3$, 
$\beta_3= 3.162 \times 10^3$, 
$\beta_4= 10^4$, 
$\beta_5= 3.162 \times 10^4$, 
$\beta_6= 5.623 \times 10^4$, 
$\beta_7= 10^5$. 
$\beta_8= 1.778 \times 10^5$, 
$\beta_9= 3.162 \times 10^5$.

We can now rewrite equation
(\ref{eq:cont-eq}) into the 
dimensionless form,
\begin{equation}
  \frac{dn_\pm}{d\xi} = 
    \pm \frac{B_{\rm cnt}}{B(\xi)}
        \sum_{i=1}^{9} [ \eta_{\rm p+}{}^i g_+^i(\xi)
                        +\eta_{\rm p-}{}^i g_-^i(\xi)],
  \label{eq:basic-3}
\end{equation}
where $\eta_{{\rm p}\pm}^i$ 
are evaluated at the central energy in each bin as
\begin{equation}
  \eta_{{\rm p}\pm}{}^i \equiv 
  \frac{1}{\omega_{\rm p}} \,
  \eta_{{\rm p}\pm} \left( \frac{\beta_{i-1}+\beta_i}{2} \right).
\end{equation}

A combination of equations (\ref{eq:basic-3}) 
gives the current conservation law,
\begin{equation}
  j_{\rm tot} \equiv n_+(\xi) + n_-(\xi) = {\rm constant \ for \ } \xi.
  \label{eq:consv}
\end{equation}
When $j_{\rm tot}=1.0$, 
the current density per unit flux tube
equals the Goldreich--Julian value, $\Omega / (2\pi)$.

\subsection{Gamma-Ray Boltzmann Equations}
\label{sec:Boltzmann_gamma}

Let us next consider the $\gamma$-ray Boltzmann equations.
The $\gamma$-rays are beamed in the direction of the
magnetic field where it was emitted.
Therefore, the propagation direction of each $\gamma$-ray photon 
does not coincide with the local magnetic field where the 
$\gamma$-ray distribution is evaluated.
However, to avoid complications, we simply assume 
that the outwardly (or inwardly) propagating $\gamma$-rays
dilate (or constrict) at the same rate with the magnetic field
(Paper VII).
Then we can write down the $\gamma$-ray Boltzmann equations into the form,
\begin{eqnarray}   
  \pm B \frac{d}{ds} \left[ \frac{1}{B} G_\pm(x,\epsilon_\gamma) \right]
     = - \frac{1}{c} \eta_{{\rm p}\pm} G_\pm(x,\epsilon_\gamma)  
       + \frac{1}{c} \eta_{\rm c} N_\pm(x),
  \label{eq:Boltz-gam}
\end{eqnarray}   
where (e.g., Rybicki, Lightman 1979)
\begin{equation}
  \eta_{\rm c} \equiv \frac{\sqrt{3}e^2 \Gamma}{h \Rc}
               \frac1{\epsilon_\gamma} 
	       F \left( \frac{\epsilon_\gamma}{\epsilon_{\rm c}} \right) ,
  \label{eq:def-etaC}
\end{equation}
\begin{equation}
  \epsilon_{\rm c} \equiv \frac1{m_{\rm e}c^2} \frac3{4\pi}
                          \frac{hc \Gamma^3}{\Rc} ,
\end{equation}
\begin{equation}
  F(s) \equiv s \int_x^\infty K_{\frac53} (t) dt ;
\end{equation}
$\Rc$ is the curvature radius of the magnetic field lines
and $K_{5/3}$ is the modified Bessel function of $5/3$ order.
The effect of the broad spectrum of curvature $\gamma$-rays
is represented by the factor $F(\epsilon_\gamma/\epsilon_{\rm c})$
in equation (\ref{eq:def-etaC}).
In the polar coordinates ($r$,$\theta$), 
the explicit expression of $\Rc$ is given by
\begin{equation}
  \frac{\Rc}{\rlc}
  = \left| 1 + \left(\frac{d\zeta}{d\xi}\right)^2 \right|^{3/2}
    \times
    \left| \frac{d^2\zeta}{d\xi^2} \right|^{-1},
  \label{eq:Rc}
\end{equation}
where
\begin{equation}
  \zeta \equiv r\cos\theta / \rlc,
\end{equation}
\begin{equation}
  \xi   \equiv r\sin\theta / \rlc.
\end{equation}
Assuming a Newtonian dipole magnetic field with inclination $\inc$,
we can write the derivatives in equation(\ref{eq:Rc}) as
\begin{equation}
  \frac{d\zeta}{d\xi}
  = \frac{ \cos\inc - (3/2)\sin\theta \sin(\theta-\inc) } 
         { \sin\inc + (3/2)\cos\theta \sin(\theta-\inc) }, 
  \label{eq:Num_Rc}
\end{equation}
\begin{equation}
  \frac{d^2\zeta}{d\xi^2}
  = \frac{1}{\rho}
    \frac{ (d\zeta/d\xi) f
           -\sin\theta\cos\inc + (3/2)\cos\theta\sin\inc } 
         { \sin\inc + (3/2)\cos\theta \sin(\theta-\inc) },
  \label{eq:Den_Rc}
\end{equation}
where
\begin{equation}
  \frac{\sin^2(\theta-\inc)}{\rho}
  = \frac{\sin^2(\theta_\ast-\inc)}{r_\ast/\rlc},
\end{equation}
\begin{equation}
  f \equiv (\cos\theta\sin\inc -\frac32 \sin\theta\cos\inc)
              \frac{d\zeta}{d\xi}
            + \frac12\cos(\theta+\inc).
\end{equation}

Integrating (\ref{eq:Boltz-gam}) in the energy intervals
[$\beta_{i-1}$, $\beta_{i}$],
we obtain
\begin{eqnarray}   
  \pm \frac{d}{d\xi} g_\pm{}^i(\xi)
     = \pm \frac{d}{d\xi}\left( \ln B \right) g_\pm{}^i
       - \eta_{{\rm p}\pm}{}^i g_\pm{}^i
       + \eta_{\rm c}^i \frac{B(\xi)}{B_{\rm cnt}} n_\pm,
  \label{eq:basic-5}
\end{eqnarray}   
where $i=1,2,\cdot\cdot\cdot,m$ ($m=9$) and 
\begin{eqnarray}
  \eta_{\rm c}^i 
  &\equiv& \frac{\sqrt{3}e^2\Gamma}{\omega_{\rm p}h \Rc}
           \int_{\beta_{i-1} / \epsilon_{\rm c}}
               ^{\beta_i     / \epsilon_{\rm c}}
            du \int_u^\infty K_{\frac53}(t)dt 
  \nonumber \\ 
  &=& 2.14 \times 10^{-8} \, \Gamma 
      \left( \frac{\Rc}{0.5\rlc} \right)^{-1}
      \sqrt{ \frac{\Omega_2}{B_5} }
  \nonumber \\
  & & \times \int_{\beta_{i-1} / \epsilon_{\rm c}}
                 ^{\beta_i     / \epsilon_{\rm c}} ds
             \int_s^\infty K_{\frac53}(t)dt .
  \label{eq:etaCi}
\end{eqnarray}

\subsection{Terminal Lorentz Factor}
\label{sec:terminal}

The most effective assumption for particle motion in the gap
arises from the fact that the velocity immediately saturates 
in the balance between 
the electric force and the radiation reaction force due to
curvature radiation.
Equating the electric force $e \vert d\Phi / dx \vert$ and the
radiation reaction force, 
we obtain the saturated Lorentz factor at each point,
\begin{equation}
  \Gamma_{\rm sat} 
   = \left( \frac{3 \Rc{}^2}{2e} 
		    \left| \frac{d\Phi}{dx} \right|
                  + 1 
     \right)^{1/4}.
  \label{eq:saturated}
\end{equation}
When $W$ is so small that a significant fraction of the 
particles are unsaturated, equation~(\ref{eq:saturated}) 
overestimates the particle's Lorentz factors.
To take account of such unsaturated motion of the particles,
we compute $\Gamma$ by 
\begin{equation}
  \frac{1}{\Gamma}
  = \sqrt{ \frac{1}{\Gamma_{\rm sat}{}^2}
          +\frac{1}{\varphi^2(\xi_2)}
         },
  \label{eq:terminal}
\end{equation}
where
\begin{equation}
  \varphi(\xi_2)= \frac{e \Phi(x_2)}
                       {m_{\rm e} c^2}
  \label{eq:maxGamma}
\end{equation}
represents the maximum attainable Lorentz factor;
the gap is considered to exists in $\xi_1 < \xi < \xi_2$
and the potential origin is chosen such that $\varphi(\xi_1)=0$.

\subsection{X-Ray Field}
\label{sec:X-ray}

To execute the integration in equation~(\ref{eq:def_etap_0}),
we must specify the X-ray field illuminating the gap.
X-ray field of a rotation-powered neutron star 
within the light cylinder
can be attributed to the following three emission processes: \\
(1) Photospheric emission from the whole surface of
a cooling neutron star
(Greenstein \& Hartke 1983; Romani 1987; Shibanov et al. 1992;
 Pavlov et al. 1994; Zavlin et al. 1995). \\
(2) Thermal emission from the neutron star's polar caps 
which are heated by the bombardment of relativistic particles
streaming back to the surface from the magnetosphere
(Kundt \& Schaaf 1993; Zavlin, Shibanov, \& Pavlov 1995;
 Gil \& Krawczyk 1996). \\
(3) Non-thermal emission from relativistic particles accelerated 
in the pulsar magnetosphere
(Ochelkov \& Usov 1980a,b; El-Gowhari \& Arponen 1972;
 Aschenbach \& Brinkmann 1975; Hardee \& Rose 1974; Daishido 1975).

The spectrum of the first component is expected to be 
expressed with a modified blackbody.
However, for simplicity, we approximate it in terms of a Plank function 
with temperature $kT_{\rm s}$, 
because the X-ray spectrum is occasionally fitted 
by a simple blackbody spectrum.
We regard a blackbody component as the first one
if its observed emitting area, $A_{\rm s}$, is
comparable with the whole surface of a neutron star,
$A_* \equiv 4 \pi r_*{}^2$,
where $r_*$ denotes the neutron star radius. 
We take account of both the pulsed and the non-pulsed 
surface blackbody emission 
as this soft blackbody component.
At distance $r$ from the center of the star, 
the X-ray density between energies $\Ex$ and $\Ex+d\Ex$ 
is given by the Planck law,
\begin{equation}
  \frac{dN_{\rm x}}{d\epsilon_{\rm x}}
  =  \frac{1}{4\pi^2} 
           \left( \frac{m_{\rm e}c^2}{c \hbar} \right)^3
           \left( \frac{A_{\rm s}}{4\pi r^2} \right)
      \frac{\Ex{}^2}
           {\exp(\Ex/\Delta_{\rm s})-1},
  \label{eq:def_Ns0}
\end{equation}
where
$\Delta_{\rm s}$ is defined by
\begin{equation}
  \Delta_{\rm s} \equiv \frac{k T_{\rm s}}{m_{\rm e}c^2};
  \label{eq:def_Ds}
\end{equation}
$kT_{\rm s}$ refers to the 
soft blackbody temperature measured by a distant observer.
Since the outer gap is located outside of the deep gravitational
potential well of the neutron star, 
the photon energy there is essentially the same as 
what a distant observer measures.

As for the second component, we regard a blackbody component as
the heated polar cap emission if its observed emitting area,
$A_{\rm h}$, is much smaller than $A_*$.
We approximate its spectrum by a Planck function.
We take account of both the pulsed and the non-pulsed 
polar cap emission 
as this hard blackbody component.
In the same manner as in the soft blackbody case, the spectrum at
a radius $r$ becomes
\begin{equation}
  \frac{dN_{\rm x}}{d\epsilon_{\rm x}}
  =  \frac{1}{4\pi^2} 
           \left( \frac{m_{\rm e}c^2}{c \hbar} \right)^3
           \left( \frac{A_{\rm h}}{4\pi r^2} \right)
      \frac{\Ex{}^2}
           {\exp(\Ex/\Delta_{\rm s})-1},
  \label{eq:def_Nh0}
\end{equation}
where
$\Delta_{\rm h}$ is related with the hard blackbody temperature,
$kT_{\rm h}$, as
\begin{equation}
  \Delta_{\rm h} \equiv \frac{k T_{\rm h}}{m_{\rm e}c^2}.
  \label{eq:def_Dh}
\end{equation}

Unlike the first and the second components, 
a power-law component is usually dominated by a nebula emission.
To get rid of the nebula emission, 
which illuminates the outer gap inefficiently,
we adopt only the pulsed components of a power-law emission
as the third component.
We describe a magnetospheric component with the power law,
\begin{equation}
  \frac{dN_{\rm pl}}{d\epsilon_{\rm x}}
    = N_{\rm pl} \epsilon_{\rm x}{}^\alpha
  \quad (\epsilon_{\rm min} < \epsilon_{\rm x} < \epsilon_{\rm max}).
  \label{eq:def-Npl}
\end{equation}
The photon index $\alpha$ is typically between $-2$ and $-1$
for a pulsed, power-law X-ray component in hard X-ray band
(e.g., Saito 1998).
We assume 
$\epsilon_{\rm min}=0.1 \mbox{keV} / 511 \mbox{keV}$ and
$\epsilon_{\rm max}=100 \mbox{keV} / 511 \mbox{keV}$
for homogeneous discussion.

\subsection{Boundary Conditions}
\label{sec:BD}

To solve the Vlasov equations
(\ref{eq:basic-1}), (\ref{eq:basic-2}), (\ref{eq:basic-3}), 
and (\ref{eq:basic-5}),
we must impose boundary conditions.
The inner boundary is defined so that 
$\Ell$ vanishes there.
Therefore, we have
\begin{equation}
  \Ell(\xi_1)=0 .
  \label{eq:BD-1}
\end{equation}
It is noteworthy that condition (\ref{eq:BD-1}) is consistent with
the stability condition at the plasma-vacuum interface 
(Krause-Polstorff \& Michel 1985a,b).
At the inner (starward) boundary
($\xi= \xi_1$), we impose (Paper VI)
\begin{equation}
  \varphi(\xi_1) = 0,
  \label{eq:BD-2}
\end{equation}
\begin{equation}
  g_+{}^i(\xi_1)=0  \quad (i=1,2,\cdot\cdot\cdot,m),
  \label{eq:BD-3}
\end{equation}
where $m=9$.
Since the particles may be created inside of the outer gap
(i.e., at $\xi<\xi_1$),
positrons may flow into the gap at $\xi=\xi_1$
as a part of the global current pattern in the magnetosphere.
We thus denote the positronic current divided by $ce$ 
per unit magnetic flux tube at $\xi=\xi_1$ as
\begin{equation}
  n_+(\xi_1)= j_1.
  \label{eq:BD-4}
\end{equation}
This yields, with the help of the charge-conservation law
(eq.~[\ref{eq:consv}]),
\begin{equation}
  n_-(\xi_1)= j_{\rm tot}-j_1.
  \label{eq:BD-5}
\end{equation}

At the outer boundary ($\xi=\xi_2$), we impose
\begin{equation}
  E_\parallel(\xi_2)=0,
  \label{eq:BD-6}
\end{equation}
\begin{equation}
  g_-{}^i(\xi_2)=0 \quad (i=1,2,\cdot\cdot\cdot,m),
  \label{eq:BD-7}
\end{equation}
\begin{equation}
  n_-(\xi_2)= j_2.
  \label{eq:BD-8}
\end{equation}

The current (divided by $ce$) created in the gap per unit flux tube
can be expressed as
\begin{equation}
  j_{\rm gap}= j_{\rm tot} -j_1 -j_2.
  \label{eq:Jgap}
\end{equation}
We adopt $j_{\rm gap}$, $j_1$, and $j_2$
as the free parameters in this paper.

We have totally $2m+6$ boundary conditions 
(\ref{eq:BD-1})--(\ref{eq:BD-8})
for $2m+4$ unknown functions
$\Phi$, $E_\parallel$,
$n_+$, $n_-$, 
$g_+{}^1$, $g_+{}^2$, $\cdot\cdot\cdot$, $g_+{}^m$, 
$g_-{}^1$, $g_-{}^2$, $\cdot\cdot\cdot$, $g_-{}^m$.
Thus two extra boundary conditions must be compensated 
by making the positions of the boundaries $\xi_1$ and $\xi_2$ be free.
The two free boundaries appear because $E_\parallel=0$ is imposed at 
{\it both} the boundaries and because $j_{\rm gap}$ is externally imposed.
In other words, the gap boundaries ($\xi_1$ and $\xi_2$),
and hence the position shifts
if $j_1$ and/or $j_2$ varies.

\section{TeV Spectra}
\label{sec:TeV_spc}

In TeV energies 
inverse Compton (IC) scatterings of infrared (IR) photons off relativistic
electrons and positrons ($\Gamma \sim 10^7$) are the process of 
$\gamma$-ray production.
We briefly discuss the IR field in \S~\ref{sec:IRfield},
and describe the intrinsic TeV emission from the gap 
in \S~\ref{sec:ICscatt} 
and the extrinsic absorption due to magnetospheric IR field 
in \S~\ref{sec:abs_TeV}.

\subsection{Infrared photon field}
\label{sec:IRfield}

Without a careful consideration of the synchrotron beaming,
it is difficult to estimate the specific intensity of the IR field.
Therefore, we simply assume that the IR field are homogeneous
and isotropic within the radius $\rlc$.
This assumption comes from the following emission scenario in the
magnetosphere:
The primary $\gamma$-rays emitted in the gap collide with X-ray
and IR photons outside of the gap to materialize as primary pairs,
which emit secondary photons in X-ray and soft $\gamma$-ray energies 
via synchrotron process.
Some portions of the secondary photons collide with each other to 
materialize as tertiary pairs in the magnetosphere.
The tertiary pairs have much lower energies compared with the secondary
pairs and emit copious tertiary IR photons via synchrotron process.
On these grounds, we may expect that the tertiary IR field 
has lost the directional information of the primary $\gamma$-rays,
compared with the secondary X-ray field.

We do not adopt a broken power-law to describe the soft photons 
from IR to X-ray energies.
This is because the adopted assumptions are different each other.
As described in \S~\ref{sec:X-ray},
we assume that the secondary, power-law X-ray field 
is homogeneous only within the radius of $r_{\rm cnt}$ around the gap
and is highly beamed along the magnetic field at the gap cener
so that the collision angles between the primary $\gamma$-rays
and the secondary X-rays in the gap are typically $W/\rlc$ radian
(eq.~[\ref{eq:def_coll_power}]).
On the other hand, we assume that the tertiary, power-law IR field 
is homogeneous and isotropic within the radius $\rlc$.

At distance $r$ from the center of the star, 
we adopt the following power-law IR spectra:
\begin{equation}
  \frac{dN_{\rm IR}}{d\epsilon_{\rm IR}}
    = N_{\rm IR} \left( \frac{d}{\rm kpc} \right)^2 
      \left(\frac{r}{\rlc}\right)^{-2}
      \epsilon_{\rm IR}{}^\alpha,
  \label{eq:IR_spctr}
\end{equation}
where $\epsilon_{\rm IR} m_{\rm e}c^2$ 
refers to the IR photon energy,
and 
$\epsilon_{\rm IR,min} < \epsilon < \epsilon_{\rm IR,max}$.
We adopt 
$\epsilon_{\rm IR,min}= 10^{-6}$ and
$\epsilon_{\rm IR,min}= 10^{-2}$;
the results do not depend on these cut-off energies very much.

\subsection{Inverse Compton Scatterings}
\label{sec:ICscatt}

When an electron or a positron is migrating 
with Lorentz factor $\Gamma \gg 1$ in an isotropic photon field,
it upscatters the soft photons to produce
the following number spectrum of $\gamma$-rays
(Blumenthal \& Gould 1970):
\begin{eqnarray}
  \frac{d^2 N}{dtd\Eg}
  &=& \frac34 \sgT \frac{c}{\Gamma^2}
      \frac{dN_{\rm IR}}{d\epsilon_{\rm IR}}
      \frac{d\epsilon_{\rm IR}}{\epsilon_{\rm IR}}
  \nonumber \\
  & & \hspace{-2.0 truecm}
      \times
      \left[ 2q \ln q +(1+2q)(1-q)
            +\frac{(Qq)^2(1-q)}{2(1+Qq)}
      \right],
  \label{eq:spc_tev}
\end{eqnarray}
where $Q \equiv 4 \epsilon_{\rm IR} \Gamma$ and 
$q \equiv \Eg / Q(\Gamma-\Eg)$;
here, $\Eg$ refers to the energy of the upscattered photons in 
$m_{\rm e}c^2$ unit.
Substituting equation~(\ref{eq:IR_spctr}),
integrating $d^2 N/dtd\Eg$ over $\epsilon_{\rm IR}$,
and multiplying the $\gamma$-ray energy ($\Eg m_{\rm e}c^2$) and
the electron number ($N_{\rm e}$) in the gap,
we obtain the flux density of the upscattered photons 
as a function of $\Eg$.

\subsection{Absorption due to pair production}
\label{sec:abs_TeV}

As mentioned in \S~\ref{sec:IRfield},
the IR field is assumed to be homogeneous and isotropic within 
radius $\rlc$.
Outside of $\rlc$, 
both the IR photon density and the collision angles decrease;
therefore, we neglect the pair production at $r>\rlc$ for simplicity. 
The optical depth then becomes
\begin{equation}
  \tau(\Eg) 
  = L \int_{\epsilon_{\rm IR,min}}^{\epsilon_{\rm IR,max}}
          \frac{dN_{\rm IR}}{d\epsilon_{\rm IR}}
          \sgP(\epsilon_{\rm IR},\Eg,\mu_{\rm c}) d\epsilon_{\rm IR},
  \label{eq:tauTeV}
\end{equation}
where $L$ refers to the path length.
For homogeneous discussion, we assume $L= 0.5 \rlc$ in this paper.
As the first order approximation, 
we adopt $90^\circ$ as the collision angles in the isotropic IR field 
to substitute $\mu_{\rm c}=0$ in equation~(\ref{eq:tauTeV}),
where $\sgP$ is defined by equation~(\ref{eq:def-sgP}).

\section{Application to Individual Pulsars}
\label{sec:appl}

In this section, we apply the theory to the nine rotation-powered
pulsars of which X-ray field at the outer gap can be deduced 
from observations.
The Crab pulsar was investigated in detail in Paper~VII;
therefore, we exclude this young pulsar in this paper.
We first describe their X-ray and infrared fields 
in the next two subsections,
and present the electric field distribution in \S~\ref{sec:Ell}
and the resultant GeV and TeV emissions from individual pulsars
in \S~\ref{sec:spctr}.

\subsection{Input X-ray Field}
\label{sec:input_Xray}

We present the observed X-ray properties of individual pulsars
in order of spin-down luminosity, $\dot{E}_{\rm rot}$ (table~1).
We assume 
$\epsilon_{\rm min}=0.1 \mbox{keV} / 511 \mbox{keV}$ and
$\epsilon_{\rm max}=100 \mbox{keV} / 511 \mbox{keV}$
for homogeneous discussion.

\begin{table*}
  \centering
    \begin{minipage}{140mm}
      \caption{Input X-ray field}
      \begin{tabular}{@{}lccccccccc@{}}
        \hline
        \hline
        pulsar	
		& distance
		& $\Omega$		& $\log_{10}\mu$
		& $kT_{\rm s}$		& $A_{\rm s}/A_*$      
		& $kT_{\rm h}$		& $A_{\rm h}/A_*$      
		& $N_{\rm pl}$		& $-\alpha$		\\
        \	
		& kpc
		& rad s${}^{-1}$	& lg(G cm${}^3$)
		& eV			&
		& eV			& 
		& cm${}^{-3}$		&			\\
        \hline
        B0540--69
		& 49.4
		& 124.7		& 31.00
		& $\ldots$	& $\ldots$
		& $\ldots$	& $\ldots$
		& $10^{14.15}$	& $2.0$				\\
        B1509--58
		& 4.40
		& 41.7		& 31.19
		& $\ldots$	& $\ldots$
		& $\ldots$	& $\ldots$
		& $10^{14.04}$	& $1.1$				\\
        J1617--5055	
		& 3.30
		& 90.6		& 30.78
		& $\ldots$	& $\ldots$
		& $\ldots$	& $\ldots$
		& $10^{12.64}$	& $1.6$				\\
        J0822--4300
		& 2.20
		& 83.4		& 30.53
		& 280		& 0.040
		& $\ldots$	& $\ldots$
		& $\ldots$	& $\ldots$			\\
        Vela	
		& 0.50
		& 61.3		& 30.53
		& 150		& 0.066
		& $\ldots$	& $\ldots$
		& $\ldots$	& $\ldots$			\\
        B1951+32
		& 2.5
		& 159		& 29.68
		& $\ldots$	& $\ldots$
		& $\ldots$	& $\ldots$
		& $10^{13.55}$	& 1.6				\\
        Geminga
		& 0.16
		& 26.5		& 30.21
		& 48		& 0.16
		& $\ldots$	& $\ldots$
		& $10^{5.00}$	& $1.6$				\\
        B1055--52
		& 1.53
		& 31.9		& 30.03
		& 68		& 7.3
		& 320		& $10^{-3.64}$
		& $\ldots$	& $\ldots$			\\
        J0437--4715
		& 0.180
		& 1092		& 26.50
		& 22		& 0.16
		& 95		& $10^{-3.28}$
		& $\ldots$	& $\ldots$			\\
        \hline
      \end{tabular}
    \end{minipage}
\end{table*}

\noindent
{\bf B0540--69} \ 
From ASCA observations in 2-10 keV band, 
its X-ray radiation is known to be well fitted by
a power-law with $\alpha=-2.0$.
The unabsorbed luminosity in this energy range is 
$1.3 \times 10^{36} \mbox{erg s}{}^{-1}$,
which leads to $N_{\rm pl}= 9.0 \times 10^9 d^2 (r_0/\rlc)^{-2}$
(Saito 1998), where $d$ is the distance to the pulsar in kpc.\\
{\bf B1509--58} \ 
From ASCA observations in 2-10 keV band, 
its pulsed emission can be fitted by
a power-law with $\alpha=-1.1$.
The unabsorbed flux in this energy range is 
$2.9 \times 10^{-11} \mbox{erg/cm}^2 / \mbox{s}$,
which leads to $N_{\rm pl}= 9.3 \times 10^{11} d^2 (r_0/\rlc)^{-2}$. \\
{\bf J1617--5055} and {\bf J0822--4300} \ 
These two pulsars have resemble parameters such as
$\Omega \sim 90 \,\mbox{rad s}^{-1}$,
$\mu \sim 10^{30.6} \mbox{G cm}^3$,
and characteristic age $\sim 8 \times 10^3$yr.
From the ASCA observations of J1617--5055 in 3.5-10 keV band, 
its pulsed emission 
subtracted the background and the steady components
can be fitted by a power-law with $\alpha=-1.6$
(Torii et al. 1998).
Adopting the distance to be 3.3 kpc (Coswell et al. 1975),
we can calculate its unabsorbed flux as
$3.1 \times 10^{-12} \mbox{erg/cm}^2 / \mbox{s}$,
which yields $N_{\rm pl}= 6.5 \times 10^{10} d^2 (r_0/\rlc)^{-2}$. 
On the other hand, 
the distance of J0822--4300 was estimated from VLA observations
as $d=2.2 \pm 0.3$kpc (Reynoso et al. 1995). 
ROSAT observations revealed that the soft X-ray
emission of this pulsar is consistent with a single-temperature
blackbody model with $kT_{\rm s}=0.28\pm 0.10$keV
and $A_{\rm s} \sim 0.04 A_* (d/2.2)^2$
(Petre, Becker, \& Winkler 1996). \\
{\bf Vela} \ 
From ROSAT observations in 0.06-2.4 keV, 
the spectrum of its point source (presumably the pulsar) emission 
is expressed by two components:
Surface blackbody component with 
$kT_{\rm s}=150$eV and $A_{\rm s}=0.066 A_* (d/0.5)^2$,
and a power-law component with $\alpha=-3.3$
($\ddot{\rm O}$gelman et al. 1993).
However, the latter component does not show pulsations;
therefore, we consider only the former component
as the X-ray field illuminating the outer gap. \\
{\bf B1951+32} \ 
From ROSAT observations in 0.1-2.4 keV,
the spectrum of its point source (presumably the pulsar) emission 
can be fitted by a single power-law component with $\alpha=-1.6$ and
intrinsic luminosity of 
$2.3 \times 10^{33} (d/2.5)^2 \mbox{erg s}^{-1}$
(Safi-Harb \& $\ddot{\rm O}$gelman 1995),
which yields 
$N_{\rm pl}=9.1 \times 10^{11} d^2 (r_0/\rlc)^{-2}$.
The extension of this power-law is consistent with the upper limit 
of the pulsed component in 2-10 keV energy band
(Saito 1998). \\
{\bf Geminga} \ 
The X-ray spectrum consists of two components:
the soft surface blackbody with $kT_{\rm s}=50$ eV and 
$A_{\rm s}= 0.22 A_* (d/0.16)^2$
and a hard power law with $\alpha= -1.6$ and
$N_{\rm pl} = 3.9 \times 10^6 d^2$ (Halpern \& Wang 1997). 
A parallax distance of 160pc was estimated from HST observations
(Caraveo et al. 1996). \\
{\bf B1055--52} \ 
Combining ROSAT and ASCA data, Greiveldinger et al. (1996)
reported that the X-ray spectrum consists of two components:
a soft blackbody with $kT_{\rm s}=68$ eV and 
$A_{\rm s}= 7.3 A_* (d/1.53)^2$
and a hard blackbody with $kT_{\rm h}=320$ eV and 
$A_{\rm h}= 2.3 \times 10^{-4} A_* (d/1.53)^2$. \\
{\bf J0437--4715} \ 
Using ROSAT and EUVE data 
(Becker \& Tr$\ddot{\rm u}$mper 1993; Halpern et at. 1996),
Zavlin and Pavlov (1998) demonstrated that both the spectra and the
light curves of its soft X-ray radiation can originate
from hot polar caps with a nonuniform temperature distribution and 
be modeled by a step-like functions having two different temperatures.
The first component is the emission from heated polar-cap core
with temperature $kT_{\rm h} = 10^{6.16}$ K measured at the
surface and with an area
$A_{\rm h} = 5.3 \times 10^{-4} A_* (d/0.180)^2$.
The second one can be interpreted as a cooler rim around the
polar cap on the neutron star surface with temperature
$kT_{\rm s} = 10^{5.53}$ K and with an area 
$A_{\rm s} = 0.16 A_* (d/0.180)^2$.
Considering the gravitational redshift factor of 0.76,
the best-fit temperatures observed at infinity become
$kT_{\rm s}= 22$ eV and $kT_{\rm h}= 95$ eV.
From parallax measurements, its distance is reported to be
$178 \pm 26$ pc (Sandhu et al. 1997).
We adopt $d=180$ pc as a representative value.

\subsection{Input Infrared Field}
\label{sec:input_IR}

We next consider the homogeneous and isotropic infrared field 
in the magnetosphere.
In addition to the references cited below,
see also Thompson et al. (1999).

{\bf B0540--69} \quad
Its de-extincted optical and soft X-ray pulsed flux densities 
can be interpolated as $F_\nu = 0.207 \nu^{-0.3}$ Jy
(Middleditch \& Pennypacker 1985). 
This line extrapolates to 0.47 mJy at 640 MHz,
which is consistent with an observed value 0.4 mJy
(Manchester et al. 1993).
We thus extrapolates the relation $F_\nu = 0.207 \nu^{-0.3}$
to the infrared energies and obtain (Paper~V)
\begin{equation}
  dN_{\rm IR}/d\epsilon_{\rm IR}
    = 1.6 \times 10^{12} d^2
      \epsilon_{\rm IR}{}^{-1.3}
  \label{eq:IR_B0540}
\end{equation}

{\bf B1509--58} \quad
The flux densities emitted from close to the neutron star
in radio (Taylor, Manchester, \& Lyne 1993), 
optical (Caraveo, Mereghetti, \& Bignami 1994),
soft X-ray (Seward et al. 1984),
and hard X-ray (Kawai et al. 1993) bands
can be fitted by a single power-law
$F_\nu = 1.36 \nu^{-0.32}$ Jy.
We thus adopt 
\begin{equation}
  dN_{\rm IR}/d\epsilon_{\rm IR}
    = 4.7 \times 10^{11} d^2
      \epsilon_{\rm IR}{}^{-1.3}
  \label{eq:IR_B1509}
\end{equation}

{\bf Vela} \quad
The flux densities emitted from close to the neutron star
in radio 
(Taylor, Manchester, \& Lyne 1993; Downs, Reichley, \& Morris 1973), 
optical (Manchester et al. 1980) bands
can be fitted by $F_\nu = 1.71 \times 10^7 \nu^{-0.91}$ Jy.
We thus adopt 
\begin{equation}
  dN_{\rm IR}/d\epsilon_{\rm IR}
    = 1.8 \times 10^{7} d^2
      \epsilon_{\rm IR}{}^{-1.9}
  \label{eq:IR_Vela}
\end{equation}

{\bf B1951+32},  \quad
The flux densities emitted from close to the neutron star
in radio (Taylor, Manchester, \& Lyne 1993)
and soft X-ray 
(Safi-Harb \& $\ddot{\rm O}$gelman, \& Finley 1995) bands
can be interpolated as $F_\nu = 32.8 \nu^{-0.49}$ Jy.
We thus adopt 
\begin{equation}
  dN_{\rm IR}/d\epsilon_{\rm IR}
    = 6.3 \times 10^{10} d^2
      \epsilon_{\rm IR}{}^{-1.5}
  \label{eq:IR_B1951}
\end{equation}


{\bf Geminga} \quad
The upper limit of the flux density in radio band 
(Taylor, Manchester, \& Lyne 1993)
and the flux density in optical band 
(Shearer et al. 1998)
gives spectral index greater (or harder) than $-0.69$.
Interpolating the infrared flux with these two frequencies,
we obtain 
\begin{equation}
  dN_{\rm IR}/d\epsilon_{\rm IR}
    = 1.9 \times 10^{7} d^2
      \epsilon_{\rm IR}{}^{-1.7},
  \label{eq:IR_Geminga}
\end{equation}
which gives a conservative upper limit of infrared photon number density
under the assumption of a single power-law interpolation.

{\bf B1055--52} \quad 
The flux densities emitted from close to the neutron star
in radio (Taylor, Manchester, \& Lyne 1993) 
and optical 
(Mignani et al. 1997) bands
can be interpolated as $F_\nu = 4.2 \times 10^5 \nu^{-0.77}$ Jy.
We thus adopt 
\begin{equation}
  dN_{\rm IR}/d\epsilon_{\rm IR}
    = 7.7 \times 10^{7} d^2
      \epsilon_{\rm IR}{}^{-1.8},
  \label{eq:IR_B1055}
\end{equation}

{\bf J1617--5055}, {\bf J0822--4300}, 
and {\bf J0437--4715} \quad
There have been no available infrarad or optical observations
for these three pulsars.
We thus simply assume that $\alpha=-1.5$ for these four pulsars
and that $\nu F_\nu = 10^9$ Jy$\cdot$Hz at 0.01 eV.
We then obtain
\begin{equation}
  dN_{\rm IR}/d\epsilon_{\rm IR}
    = 2.5 \times 10^{7} d^2
      \epsilon_{\rm IR}{}^{-1.5}
  \label{eq:IR_J1617}
\end{equation}
for J1617--5055, 
\begin{equation}
  dN_{\rm IR}/d\epsilon_{\rm IR}
    = 2.1 \times 10^{7} d^2
      \epsilon_{\rm IR}{}^{-1.5}
  \label{eq:IR_J0822}
\end{equation}
for J0822--4300, 
and
\begin{equation}
  dN_{\rm IR}/d\epsilon_{\rm IR}
    = 3.6 \times 10^{13} d^2
      \epsilon_{\rm IR}{}^{-1.5}
  \label{eq:IR_J0437}
\end{equation}
for J0437--4715.

\subsection{Electric Field Structure}
\label{sec:Ell}

To examine the behavior of the solutions in a wide parameter space,
we consider the following four cases:
$(j_1,j_2)=(0,0)$, $(0.3,0)$, $(0.6,0)$, and $(0,0.3)$. 
In what follows, we denote them as 
cases~1, 2, 3, and 4, respectively.
In case~1, particles flow into the gap at
neither the inner nor the outer boundary.
In case~2 (or case~4), 
the positronic (or electronic) current flowing into the gap
per unit flux tube at the outer (or inner) boundary is $30\%$ of 
the typical Goldreich--Julian value, $\Omega /2\pi$.

In case~1, for $\inc=45^\circ$, 
the solution for B1509--58 disappears if $j_{\rm gap}>0.0825$
and that for B1055--52 disappears     if $j_{\rm gap}>0.0152$.
This is because the electric field distribution forms a \lq brim'
(e.g., fig.~2 in Hirotani \& Okamoto 1998)
as $j_{\rm gap}$ approaches a certain upper limit,
typically several percent of $\Omega /2\pi$.
On these grounds, we fix $j_{\rm gap}= 0.01$ for all the four cases.
We consider B1509--58, B1055--52, and J0437--4715
to describe $\Ell$ distributions.

\subsubsection{B1509--58, a young pulsar}
\label{sec:Ell_1509}

Since the Crab pulsar was investigated in Paper~VII,
we consider here B1509--58 as a typical example of young pulsars.
The results of $\Ell(s)$ for $\inc=45^\circ$ is presented in
figure~\ref{fig:Ell_1509_45}.  
The solid, dashed, dash-dotted, and dotted lines correspond to the
cases 1, 2, 3, and 4, respectively.
The abscissa designates the distance along the last-open field line
and covers the range from the neutron star surface ($s=0$)
to the position where the disance equals 
$s= 1.2 \times \rlc= 9.34 \times 10^6$~m.

The solid line shows that the gap is located around the null surface,
if no currents penetrate into the gap at neither of the boundaries.
The gap slightly extends outwards, 
because the pair production mean free path becomes small 
at the outer region of the gap,
owing to the diluted X-ray field there.
It is noteworthy that the potential drop becomes the largest
when $j_1=0$ and $j_2=0$ (i.e., case~1).

The dashed and dash-dotted lines indicate three points we may notice.
For one thing, 
the gap shifts outwards as $j_1$ increases.
This conclusion is consistent with the results obtained analytically by
Shibata and Hirotani (2000).
What is more, the maximum value of $\Ell$ decreases 
as the gap shifts outwards.
This is because the decreased $\rhoGJ$
at larger distance reduces $\vert d\Ell/ds \vert$ 
(eq.[\ref{eq:Poisson_1}]).
One final point is that $W$ increases as the gap shifts outwards.
This is because the pair production mean free path increases,
owing to the diluted X-ray field at large radii.
The potential drop in the gap remains almost unchanged
between cases~2 and 3,
because the decrease of maximum $\Ell$ and the increase of $W$
cancel each other.

On the other hand, when $j_2$ increases,
the gap shifts inwards.
When the gap is located on the half way from the null surface to the star
(case~4, the dotted line),
the potential drop becomes only $3\%$ of that in case~1.
A physical interpretation on this point
is given in \S~\ref{sec:closure}.
 
\subsubsection{PSR B1055--52, a middle-aged pulsar}
\label{sec:Ell_1055}

\begin{figure} 
\centerline{ \epsfxsize=8.5cm \epsfbox[200 20 500 250]{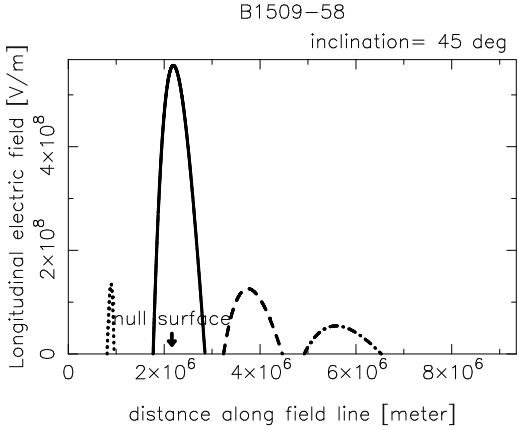} } 
\caption{\label{fig:Ell_1509_45} 
Distribution of the acceleration field in the magnetosphere of B1509--58,
a young pulsar.
The inclination angle is $45^\circ$.
The abscissa is the distance from the star surface along the
last-open fieldline.
The solid, dashed, dash-dotted, and dotted lines correspond to the
cases 1, 2, 3, and 4, respectively (see text).
        }
\end{figure}

\begin{figure} 
\centerline{ \epsfxsize=8.5cm \epsfbox[200 20 500 250]{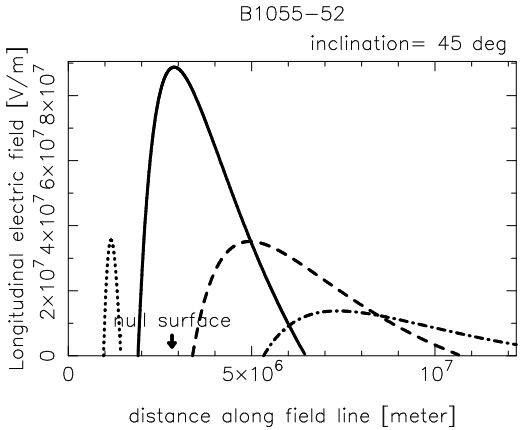} } 
\caption{\label{fig:Ell_1055_45} 
Distribution of the acceleration field for B1055--52,
a middle-aged pulsar.
The inclination angle is $45^\circ$.
The abscissa, ordinate, and the lines are the same as figure~\figtwo.
        }
\end{figure}

We next consider PSR B1055--52 as an example of middle-aged pulsars.
The $\Ell$ distribution for $\inc=45^\circ$ is presented in
figure~\ref{fig:Ell_1055_45}.  
The abscissa, ordinate, and the lines and the same as 
figure~\ref{fig:Ell_1509_45}.

Comparing with figure~\ref{fig:Ell_1509_45},
we can understand that the gap is more extended for B1055--52.
This is because the X-ray field,
which is supplied by the surface emission,
becomes less dense for middle-aged pulsars. 
The weak X-ray field enlarges the pair-production mean free path,
and hence the gap width.  
The strength of $\Ell$ decreases compared with young pulsars,
because local $\rhoGJ$ is small due to their large $\rlc$,
or equivalently, due to their small $\Omega$.

The position of the gap behaves in the same way as B1509--58.
That is, it shifts towards the light cylinder (or the star surface) 
when $j_1$ (or $j_2$) increases.

\subsubsection{J0437--4715, a millisecond pulsar}
\label{sec:Ell_0437}

\begin{figure} 
\centerline{ \epsfxsize=8.5cm \epsfbox[200 20 500 250]{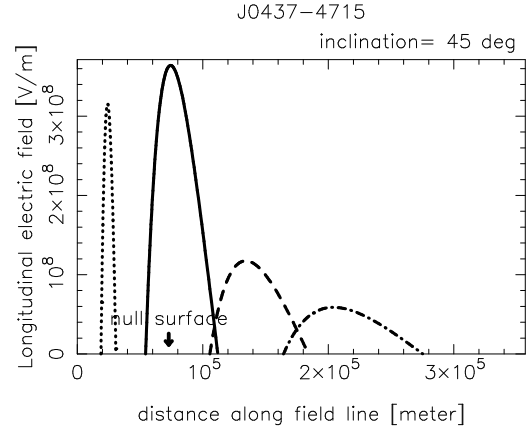} } 
\caption{\label{fig:Ell_0437_45} 
Distribution of the acceleration field for J0437--4715,
a millisecond pulsar.
The inclination angle is $45^\circ$.
The abscissa, ordinate, and the lines are the same as figure~\figtwo.
        }
\end{figure}

We finally consider J0437--4715, a millisecond pulsar.
The $\Ell$ distribution for $\inc=45^\circ$ is presented
in figure~\ref{fig:Ell_0437_45}.
Even though the magnetic moment is small,
the strength of $\Ell$ is comparable with the young pulsar B1509--58.
This is because the magnetic field in the outer magnetosphere 
is comparable with those of young pulsars, 
owing to its fast rotation ($\Omega= 1092 \mbox{rad s}^{-1}$),
which reduces $\rlc$ to $2.74 \times 10^5 \mbox{m}$.
In addition, $W$ is much less than B1055--52, a middle-aged pulsar
and relatively close to B1509--69, a young pulsar.
This is because the shrunk light cylinder of millisecond pulsars
results in a dense X-ray field illuminating the gap;
the dense X-ray field in turn reduces the pair-production mean free path,
and hence $W$.
This result is consistent with the semi-analytical prediction for 
the millisecond pulsar B1821--24 (Paper~V), 
of which X-ray field is dominated by a power-law component.
In short, the electrodynamic structures of the gap
are similar between millisecond pulsars and young pulsars.

\subsection{Gamma-ray Spectra}
\label{sec:spctr}

The spectra of outwardly (or inwardly) propagating $\gamma$-rays
in GeV energies,
are readily computed from 
$g_+{}^i(\xi_2)$ (or $g_-{}^i(\xi_1)$).
The TeV spectra, on the other hand, 
can be obtained by the method described in \S~\ref{sec:TeV_spc}.
Except for B0540--69,
the optical depth (eq.[\ref{eq:tauTeV}]) is much less than unity;
therefore, the effect of absorption due to TeV-eV collisions
in the magnetosphere, can be neglected 
for the other eight pulsars considered in this paper.

\begin{figure} 
\centerline{ \epsfxsize=8.5cm \epsfbox[200 20 500 250]{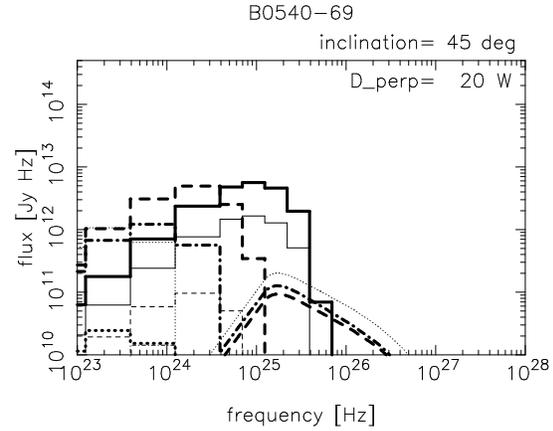}}
\caption{\label{fig:Spc_0540_45} 
Pulsed gamma-ray spectra from B0540--69 magnetosphere 
when $\inc=45^\circ$.
The thick and thin lines denote the spectra of 
outwardly and inwardly propagating $\gamma$-rays, respectively.
The solid, dashed, dash-dotted, and dotted lines correspond to the
cases 1, 2, 3, and 4, respectively (see text).
        }
\end{figure}

We present the combined GeV--TeV spectra for 
B0540--69, B1509--58, J1617--5055, J0822--4300, 
Vela, B1951+32, Geminga,
B1055--52, and J0437--4715. 
in figures~\ref{fig:Spc_0540_45}--\ref{fig:Spc_0437_45},
multiplying the same cross sectional area for the
GeV and TeV emissions for each pulsar.
Both the inclination angle and $D_\perp$ 
are indicated at the top right corner.
The outward (or inward) $\gamma$-ray flux is depicted
by thick (or thin) lines.
The observed pulsed fluxes and their upper limits
are indicated by 
open circles (EGRET),
open squares (Whipple),
open triangles (Durham group).
The upper limits of the stationary fluxes are obtained by CANGAROO group
and are denoted by filled circles.
To compare the results with the same $\inc$,
we present the case of $\inc=45^\circ$ for all the pulsars.
For Vela, B1951+32, Geminga, and B1055--52,
we also present the case of another $\inc$ to compare 
the expected spectra with EGRET observations.


\begin{figure} 
\centerline{ \epsfxsize=8.5cm \epsfbox[200 20 500 250]{Spc_1509_45.ps} } 
\caption{\label{fig:Spc_1509_45} 
Pulsed gamma-ray spectra from B1509--58 when $\inc=45^\circ$.
The lines correspond to the same cases as in \figfive.
        }
\end{figure} 


\begin{figure} 
\centerline{ \epsfxsize=8.5cm \epsfbox[200 20 500 250]{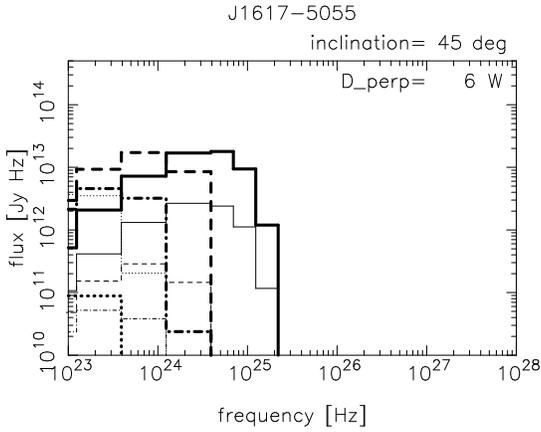} } 
\caption{\label{fig:Spc_1617_45} 
Pulsed gamma-ray spectra PSR J1617--30 when $\inc=45^\circ$.
The lines correspond to the same cases as in \figfive.
        }
\end{figure} 

\begin{figure} 
\centerline{ \epsfxsize=8.5cm \epsfbox[200 20 500 250]{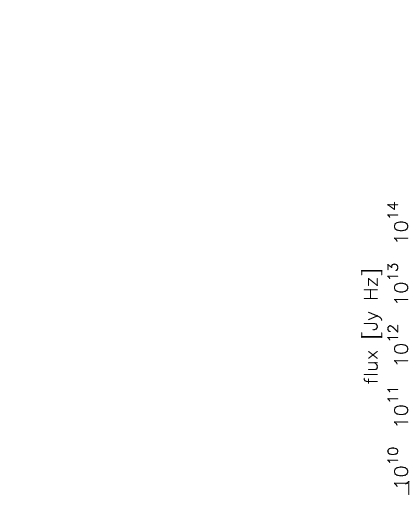} } 
\caption{\label{fig:Spc_0822_45} 
Pulsed gamma-ray spectra for the Vela pulsar when $\inc=45^\circ$.
The lines correspond to the same cases as in \figfive.
        }
\end{figure} 


\begin{figure} 
\centerline{ \epsfxsize=8.5cm \epsfbox[200 20 500 250]{Spc_Vela_45.ps} } 
\caption{\label{fig:Spc_Vela_45} 
Pulsed gamma-ray spectra for the Vela pulsar when $\inc=45^\circ$.
The lines correspond to the same cases as in \figfive.
        }
\end{figure} 

\begin{figure} 
\centerline{ \epsfxsize=8.5cm \epsfbox[200 20 500 250]{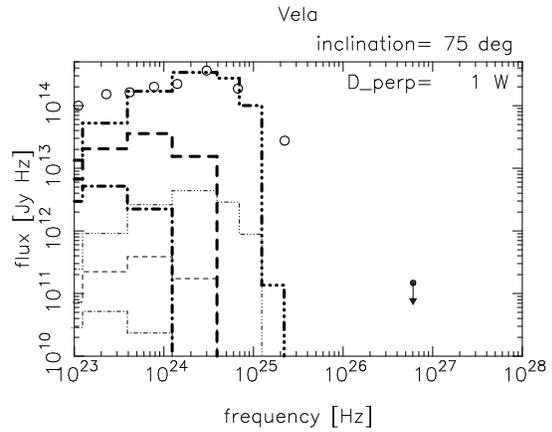} } 
\caption{\label{fig:Spc_Vela_75} 
Pulsed gamma-ray spectra for the Vela pulsar when $\inc=75^\circ$.
The dashed and dotted lines correspond to the same cases as in \figfive,
while the dash-dot-dot-dotted lines to 
$(j_{\rm gap},j_1,j_2)=(0.01,0.07,0)$. 
        }
\end{figure} 


\begin{figure} 
\centerline{ \epsfxsize=8.5cm \epsfbox[200 20 500 250]{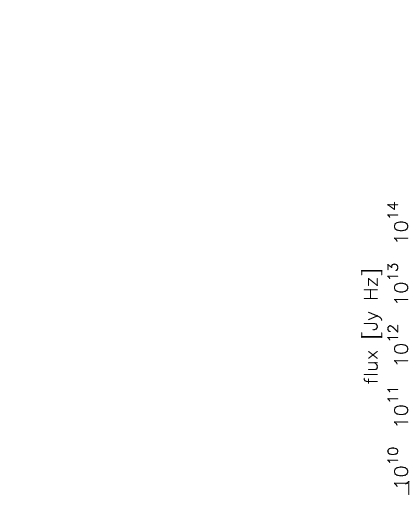} } 
\caption{\label{fig:Spc_1951_45} 
Pulsed gamma-ray spectra from B1951+32 when $\inc=45^\circ$.
The lines correspond to the same cases as in \figfive.
        }
\end{figure} 

\begin{figure} 
\centerline{ \epsfxsize=8.5cm \epsfbox[200 20 500 250]{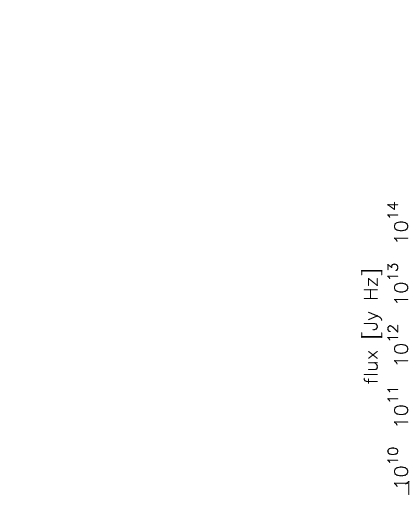} } 
\caption{\label{fig:Spc_1951_75} 
Pulsed gamma-ray spectra from B1951+32 when $\inc=75^\circ$.
The lines correspond to the same cases as in \figfive.
        }
\end{figure} 


\begin{figure} 
\centerline{ \epsfxsize=8.5cm \epsfbox[200 20 500 250]{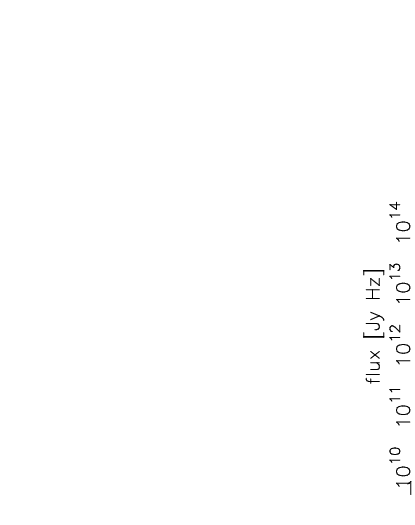} } 
\caption{\label{fig:Spc_Gemi_45} 
Pulsed gamma-ray spectra for the Geminga pulsar when $\inc=45^\circ$.
The lines correspond to the same cases as in \figfive.
        }
\end{figure} 

\begin{figure} 
\centerline{ \epsfxsize=8.5cm \epsfbox[200 20 500 250]{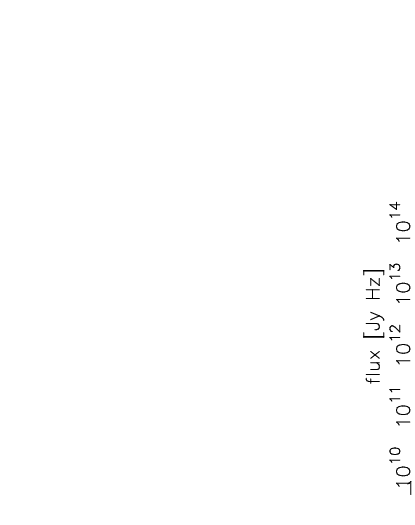} } 
\caption{\label{fig:Spc_Gemi_60} 
Pulsed gamma-ray spectra for the Geminga pulsar when $\inc=45^\circ$.
The lines correspond to the same cases as in \figfive.
        }
\end{figure} 

\begin{figure} 
\centerline{ \epsfxsize=8.5cm \epsfbox[200 20 500 250]{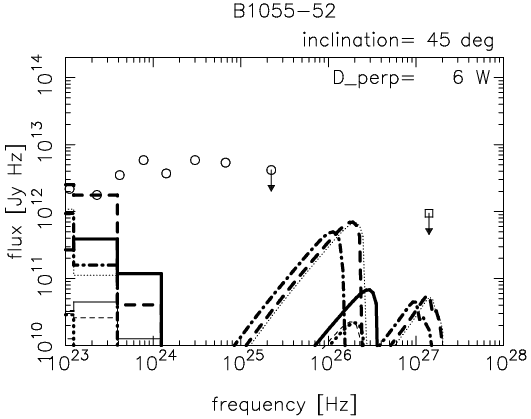} } 
\caption{\label{fig:Spc_1055_45} 
Pulsed gamma-ray spectra from B1055--52 when $\inc=45^\circ$.
The lines correspond to the same cases as in \figfive.
        }
\end{figure} 

\begin{figure} 
\centerline{ \epsfxsize=8.5cm \epsfbox[200 20 500 250]{Spc_1055_75.ps} } 
\caption{\label{fig:Spc_1055_75} 
Pulsed gamma-ray spectra from B1055--52 when $\inc=75^\circ$.
The lines correspond to the same cases as in \figfive.
        }
\end{figure} 


\begin{figure} 
\centerline{ \epsfxsize=8.5cm \epsfbox[200 20 500 250]{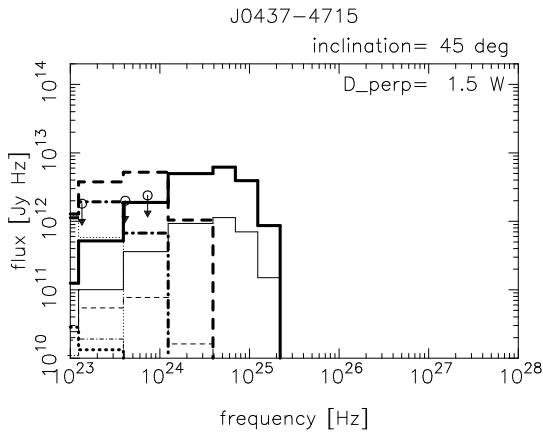} } 
\caption{\label{fig:Spc_0437_45} 
Pulsed gamma-ray spectra for J0437--4715 when $\inc=45^\circ$.
The lines correspond to the same cases as in \figfive.
        }
\end{figure} 

\subsubsection{General Features}

We first consider the general features of the $\gamma$-ray spectra.
The GeV luminosity is the largest and hardest for case~1 
(\S~\ref{sec:Ell}),
because both the potential drop and the maximum of $\Ell$ is the largest.
(see figs.~\ref{fig:Ell_1509_45}--\ref{fig:Ell_0437_45}).
It should be noted that the larger $\inc$ is,
the harder the GeV spectrum becomes;
this can be understood if we compare 
figures~\ref{fig:Spc_1951_45} and \ref{fig:Spc_1951_75}, for instance.
It should be noted that the same conclusion was derived independently
for a vacuum gap in Paper~V (e.g., $E_{\rm c}$ in table~3), 
in which the \lq gap closure' condition was considered
in stead of solving the stationary Vlasov equations (\S~\ref{sec:basicEQ}).
Physically speaking, this is because the distance of the gap
from the star decreases with increasing $\inc$ for the same set of
$(j_{\rm gap},j_1,j_2)$.
At small radii, the strong magnetic field enlarges $\rhoGJ$,
and hence $\vert d\Ell/ds\vert$,
which results in a large $\Ell$ at the gap center,
especially when the gap is nearly vacuum
(as the solid lines indicate in 
 figs.~\ref{fig:Spc_1951_45} and \ref{fig:Spc_1951_75}).

We can also understand that the TeV fluxes are generally kept 
below the observational upper limits for appropriate GeV fluxes,
or equivalently, appropriate cross sectional area, $D_\perp{}^2$.
Therefore, the problem of the excessive TeV flux
does not arise for reasonable IR densities of individual pulsars.
In the next subsection, we consider the expected spectra 
source by source.

\subsubsection{Comparison with EGRET observations}

First, we consider the Vela pulsar.
It follows from the figures~\ref{fig:Spc_Vela_45} and \ref{fig:Spc_Vela_75}
that the pulsed GeV spectrum obtained by EGRET 
(Kanbach et al. 1994; Fierro et al. 1998)
is impossible to be explained if the gap is located well inside of the 
null surface, as the dotted lines indicate.
A large inclination angle and small current inflows
are preferable to explain the observed flux around $10^{25}$~Hz.
For example, we present the case for $j_1=0.07$ and $j_2=0$ 
as the (thick and thin) dash-dot-dot-dotted lines 
in figure~\ref{fig:Spc_Vela_75}. 
For $\inc>45^\circ$,
expected TeV fluxes are well below the observed upper limit
(Yoshikoshi et al. 1997).

Secondly, we consider B1951+32.
It follows again that the observed GeV flux around $10^{25}$~Hz 
(Ramanamurthy et al. 1995) is impossible 
to be emitted if the gap is located well inside of the null surface,
as the dotted lines indicate.
Although the expected TeV flux is much less than the observed upper 
limits around $10^{26}$~Hz (Srinivanan et al. 1997),
it may be detectable around $\nu \sim 10^{27}$~Hz
with a future ground-based telescopes,
if $\inc$ is as large as $75^\circ$.

Thirdly, we consider the Geminga pulsar.
For this middle-aged pulsar, 
only the dotted lines are depicted
in figures~\ref{fig:Spc_Gemi_45} and \ref{fig:Spc_Gemi_60}, 
because there exist no solutions for cases~1, 2, and 3. 
That is, for $\inc=45^\circ$ (or $60^\circ$), 
solutions exist only when $j_2>0.073$ (or $j_2>0.017$),
provided that $j_{\rm gap}= 0.01$ and $j_1=0$ hold.
This is because its X-ray field is so weak that the gap extends
towards the light cylinder and vanishes $d\Ell/ds$ at $\xi= \xi_2$
(i.e., forms a \lq brim')
at the critical value of $j_2$ ($=0.073$ for $\inc=45^\circ$, say).
Since the solutions exist in a limited region of the parameter space,
the observed GeV spectrum above $10^{24}$~Hz 
(Mayer-Hasselwander et al. 1994)
are impossible to be explained by the
present theory.
The TeV flux is expected to be much less than the observed upper limits
(for pulsed upper limits, see Akerlof et al. 1993;
 for steady flux and/or its upper limits, see
 Weekes and Helmken 1977, Kaul et al. 1985, 
 Vishwanath et al. 1993, Bowden et al. 1993).

Fourthly, we consider B1055--52.
For this middle-aged pulsar, solutions exist in a wide parameter range,
unlike the Geminga pulsar.
However, we obtain too soft GeV spectra to fit the observations
(Thompson et al. 1999).
The difficulties in the application to these two middle-aged pulsars 
will be discussed in \S~\ref{sec:difficult}.
 
\begin{figure} 
\centerline{ \epsfxsize=8.5cm \epsfbox[200 20 500 250]{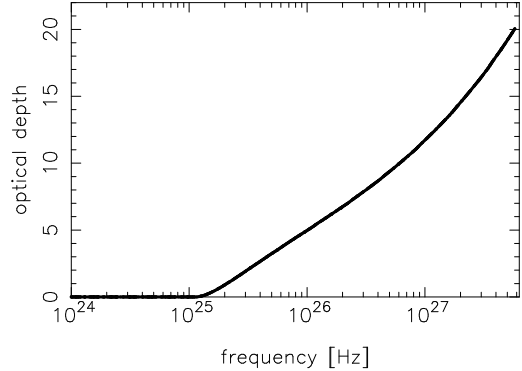}}
\caption{\label{fig:Tau_0540_30} 
Pair production optical depth for a TeV photon to be absorbed
in the homogeneous, isotropic IR field in the 
magnetosphere of pulsar B0540--69.
The abscissa designates the $\gamma$-ray frequency.
        }
\end{figure}

Finally, we consider the other pulsars of which GeV fluxes
were not detected by EGRET.
For B0540--69, the basic properties are common with the
Crab pulsar (Paper~VII) as follows:\\
\ (1) \ 
$W \ll \rlc$ holds because of its large X-ray density.\\
\ (2) \ 
Intrinsic TeV flux is comparable or even somewhat greater than the GeV 
flux but is significantly absorbed by its dense, magnetospheric IR field.
For example, in case~1, the intrinsic TeV flux before absorption 
attains $1.7 \times 10^{13}$~Jy~Hz at $1.4 \times 10^{27}$~Hz,
which is about two times greater than the maximum of the GeV flux.
In figure~\ref{fig:Tau_0540_30}, we present the absorption optical depth
(eq.~[\ref{eq:tauTeV}]).
The resultant, absorbed TeV spectra are depited in 
figure~\ref{fig:Spc_0540_45}, together with the GeV spectra.\\
\ (3) \ 
The curvature spectra are relatively hard
(fig.~\ref{fig:Spc_0540_45}),
because its strong magnetic field at the gap increases $\Ell$, 
although its $W$ is small.

For B1509--58, its relatively less dense IR field
(compared with Crab and B0540--69) cannot absorb
enough TeV flux in the magnetosphere.
As a result, the expected TeV flux becomes comparable with 
the observational upper limit while the GeV flux is well below the
observational upper limits (fig.~\ref{fig:Spc_1509_45}).
As figures~\ref{fig:Spc_1951_45} and \ref{fig:Spc_1951_75} indicate,
the TeV flux does not increase so rapidly as GeV one 
with increasing $\inc$. 
We thus conclude that $\inc$ should be comparable with or
greater than $45^\circ$ so that
the GeV and TeV fluxes may be consistent with the observations.
(Note that we can adjust the cross sectional area to decrease or increase
the GeV and TeV fluxes at the same rate.)

In the case of J1617--5055,
its GeV spectrum (fig.~\ref{fig:Spc_1617_45})
is softer than Crab (Paper~VII) and B0540--69,
because its X-ray field is less dense compared with these two
young pulsars.
Its TeV flux is small because of its small IR field assumed
(\S~\ref{sec:IRfield}).

The GeV spectrum of J0822--4300 (fig.~\ref{fig:Spc_0822_45}) 
is softer than that of Vela,
although the X-ray field more or less resembles each other.
This is because the former's high-temperature surface emission
provides a denser X-ray field, 
which shrinks the pair-production mean free path, 
and hence $W$ to reduce $\Ell$,
compared with Vela.

For J0437--4715, the GeV flux is large (fig.~\ref{fig:Spc_0437_45}),
although we assume a small cross sectional area, $D_\perp{}^2= (1.5W)^2$.
This large Gev flux comes from the fact that millisecond pulsars
have similar electrodynamic structure with young pulsars.
Compared with the EGRET upper limit of pulsed emissions (Thompson 2000),
we can expect that this milli-second pulsar is an attractive
target for the next-generation GeV telescopes.

\section{Discussion}
\label{sec:discussion}

In summary, we have developed a one-dimensional model for 
an outer-gap accelerator in the magnetosphere of a rotating neutron star.
When the positrons (or electrons) penetrate into the gap
from the inner (or outer) boundary,
the gap shifts outwards (or inwards).
Applying the theory to Vela and B1951+32,
we find that the GeV spectrum softens significantly
and cannot explain the EGRET observations,
if the gap is located well inside of the null surface.
We can therefore conclude the gap should be located near to or 
outside of the null surface for these two pulsars.
We also apply the theory to 
B0540--69, B1509--58, J1617--5055, J0822--4300, and J0437--4715
and find that 
the TeV fluxes are undetectable by current ground-based
telescopes for moderate cross sectional areas of the gap
for all the seven pulsars mentioned just above,
if we estimate their magnetospheric IR field by interpolating 
radio and optical pulsed fluxes.
For B0540--69, its intrinsically large TeV flux is absorbed 
by its dense IR field to become undetectable;
this conclusion is analogous with what was obtained in Paper~VII
for the Crab pulsar.

\subsection{Difficulties in the Application to Middle-aged Pulsars}
\label{sec:difficult}

We briefly discuss the difficulties in the application to middle-aged
pulsars.
As figures~\ref{fig:Spc_Gemi_45}--\ref{fig:Spc_1055_45} indicate,
the fluxes from Geminga and B1055--52
above $10^{24}$~Hz are quite insufficient
compared with EGRET observations.
There are three reasons for this problem.
First, special relativistic effects becomes important
for the extented gaps of middle-aged pulsars,
of which X-ray field is weak.
For example, in equation~(\ref{eq:Poisson_1}),
we neglected the terms containing higher orders in 
$(\Omega \varpi / c)^2$.
Secondly, the global geometry of the magnetic field lines becomes 
important as as the gap extends.
That is, the rectilinear approximation (\S~\ref{sec:Poisson})
cannot be applied to these two middle-aged pulsars.
Thirdly, their large $\Ell$, which is a good fraction of the
trans-field electric field, prevents particles from corotation
(eq.[\ref{eq:toroidal}]).
On these grounds, applications of the present theory
to middle-aged pulsars do not provide good agreement
with observations.

\subsection{Gap Closure Condition}
\label{sec:closure}

In \S~\ref{sec:Ell_1509},
we pointed out that the potential drop decreases significantly
as the gap shifts inwards from the null surface.
In this subsection,
we interpret this behavior from the gap closure condition
investigated in Papers~IV and V.
In a stationary gap, the pair production optical depth,
$W/\lambda_{\rm p}$, equals the ratio 
$N_\gamma(j_{\rm gap}$/$j_{\rm tot})$,
where $\lambda_{\rm p}$ and $N_\gamma$ refer to 
the pair production mean free path
and the number of $\gamma$-rays emitted by a single particle, respectively.
We thus obtain the following gap closure condition:
\begin{equation}
  W = \lambda_{\rm p} N_\gamma \frac{j_{\rm gap}}{j_{\rm tot}}.
  \label{eq:closure}
\end{equation}
This condition is automatically satisfied by the stationarity of the
Boltzmann equations.
For example, 
consider the case when the external current is ten times greater than 
the created one in the gap.
In this case, the expectation value for the $\gamma$-rays
that are emitted from a single particle in the gap,
should be one tenth so that a stationary pair production cascade may be
maintained, because not only the produced particles but also the
externally injected particles emit $\gamma$-rays in the same way.

As the gap shifts inwards,
the X-ray density increases to reduce $\lambda_{\rm p}$.
At the same time, 
the ratio $j_{\rm gap}$/$j_{\rm tot}$ decreases
as $j_2$ increases. 
As a result, $W$ decreases very rapidly with increasing $j_2$.
Owing to the rapidly decreasing $W$, 
the integrated potential drop significantly decreases,
although the local $\rhoGJ$, 
and hence $\vert d\Ell/ds \vert$ increases inwards.

If the gap shifts outwards, on the other hand,
the increase of $\lambda_{\rm p}$ and 
the decrease of $j_{\rm gap}/j_{\rm tot}$ 
partially cancel each other.
As a result, the potential drop decreases only slightly with 
increasing $j_1$.

\subsection{Comparison with ZC97}
\label{sec:ZC97}

It is worth comparing the present method with ZC97,
who onsidered that the gap size is limited so that the emitted
$\gamma$-rays may have energies just above the pair production 
threshold in collisions with the surface X-rays that are
due to the bombardment of produced particles in the gap.
Their method qualitatively agrees with the current gap closure condition
(eq.~[\ref{eq:closure}]),
provided $j_{\rm gap}/j_{\rm tot} \sim 1$ (see \S~5.8 in Paper~V);
case~1 ($j_1=j_2=0$) corresponds to this case.
However, as $j_2$ increases, the factor $(j_{\rm gap}/j_{\rm tot})$
in equation~(\ref{eq:closure}) becomes small to reduce $W$.
In other words, the closure condition adopted by ZC97 
would become inconsistent with stationary Vlasov equations (\S~2),
if we were to overextraporate it to the case when $j_2 \sim 1$.
In the same manner, 
we cannot overextraporate the gap-closure condition of ZC97
to the case when $j_1$ increases and hence the gap shifts 
towards the light cylinder (as cases~2 and 3),
because $j_{\rm gap}/j_{\rm tot}$ decreases substantially.

\subsection{Inverse Compton Scatterings in Young Pulsar Magnetosphere}
\label{sec:ICyoung}

Although the dense IR field of young pulsars (B0540--69, B1509--58)
results in a relatively large intrinsic TeV emission 
compared with other pulsars,
the intrinsic TeV flux is not more than the GeV flux,
provided $\inc>45^\circ$.
Therefore, the radiation reaction force is caused primarily by
the curvature process rather than the inverse Compton scatterings,
as long as $j_2$ is less than $0.3$.
However, if $j_2$ exceeds $0.3$, 
we cannot in general rule out the possibility of
the case when the radiation reaction is caused by IC scatterings.
This is because the dense X-ray field will suppress the particle Lorentz 
factors (Paper~II) below $10^7$,
and because such less-energetic particles scatter 
copious IR photons into lower $\gamma$-ray energies
with large cross sections ($\sim \sgT$).
As a result, the GeV emissions may be mainly due to IC scattering
in this case.
Moreover, since the magnetic field becomes very strong close to the star,
not only the synchro-curvature process (ZC97),
but also a magnetic pair production becomes important. 
There is room for further investigation of the case when 
an \lq outer' gap is located close to the polar cap.

\section*{Acknowledgments}

One of the authors (K. H.) wishes to express his gratitude to
Drs. Y. Saito and A. Harding for valuable advice. 
He also thanks the Astronomical Data Analysis Center of
National Astronomical Observatory, Japan for the use of workstations.


 \label{lastpage}

\end{document} 

\bye